\documentclass[10pt]{article}

\usepackage{amsmath}
\usepackage{amssymb}

\usepackage{graphicx}

\usepackage{cite}

\usepackage{color} 

\usepackage[labelfont=bf,labelsep=period,justification=raggedright]{caption}

\makeatletter
\renewcommand{\@biblabel}[1]{\quad#1.}
\makeatother

\date{}

\begin{document}

\begin{flushleft}
{\Large
\textbf{Categorical and Geographical Separation in Science}
}
\\
Julian Sienkiewicz$^{a,\ast}$, 
Krzysztof Soja$^{a}$, 
Janusz A. Ho{\l}yst$^{a}$ and
Peter M. A. Sloot$^{b,c,d}$
\\
\bf{a} Faculty of Physics, Center of Excellence for Complex Systems Research, Warsaw University of Technology, Warsaw, Poland
\\
\bf{b} Computational Science, University of Amsterdam, Amsterdam, The Netherlands
\\
\bf{c} National Research University of Information Technologies, Mechanics and Optics (ITMO), Saint Petersburg, Russia
\\
\bf{d} Nanyang Technological University, Singapore, Singapore
\\
$\ast$ E-mail: julas@if.pw.edu.pl
\end{flushleft}

\section*{Abstract}
We perform the analysis of scientific collaboration at the level of universities. The scope of this study is to answer two fundamental questions: (i) can one indicate a category (i.e., a scientific discipline) that has the greatest impact on the rank of the university and (ii) do the best universities collaborate with the best ones only? Using two university ranking lists (ARWU and QS) as well as data from the Science Citation Index we show how the number of publications in certain categories correlates with the university rank. Moreover, using complex networks analysis, we give hints that the scientific collaboration is highly embedded in the physical space and the number of common papers decays with the distance between them. We also show the strength of the ties between universities is proportional to product of their total number of publications.

\section*{Introduction}
The 20th century is well known for its critical works of Kuhn, Popper, Lakatos and Feyerabend that tried to build models of how the science should work or to show how it does in fact work. In the same, owing to the entrance into era of overwhelming information it was possible to tackle this problem quantitatively \cite{merton,king}, pointing out specific phenomena observed in science. Several studies are bound to answer such questions like "How to measure who the best scientist is?" \cite{hirsh,rad1,rad2,petersen,rad3,mz,fronczak} or try to simulate the process of paradigms shifts \cite{bornholdt,kondratiuk}. In this study, we make use of complex networks tools \cite{ba,chmiel} to show how this issue is resolved at the level of scientific institutions (i.e., universities), to be more specific (i) what is the correlation between university rank and the number of papers in a specific discipline and (ii) what are the components of the scientific collaborations.

\section*{Results}

In order to estimate the correlations between university rankings and scientific productivity we had to identify two different sources of data: (i) first devoted to the university ranking with at least 10 years activity , (ii) second connected to actual bibliographic information, in particular complying with the following rules: (1) allowing to view categories of publications,(2) allowing to view address of the publication, (3) allowing to view year of publication. The lists of top hundred universities were downloaded from two services: Academic Ranking of World Universities\footnote{http://www.arwu.org} - later referred to as ARWU and QS World University Ranking\footnote{http://www.topuniversities.com} - later referred to as QS. The rationale behind choosing two rankings that follow different rules was to check the robustness of the performed analysis. After preliminary analysis, we have chosen the service Web of Science as a data source for obtaining the information on citations. For one institution the average number of publications ranges between few to dozens of thousands of publications. As a result each university has two tables containing the following fields: (i) published (date of publication),(ii) ID (reference to the second table),(iii) subject category (category of publications),(iv) language. The key information used in this report is the subject category of the published paper (we will refer to it later as to simply {\it category})

We start the analysis with the estimation of correlation reflecting the dependence of the number of papers a university has published on its rank in the list. To be more precise, for each of 180 categories we build a 100 by 2 matrix, where the first row gives ranks $r_i$ of all universities in this category and the second one gathers the number of papers $n_i$ published in this category by the given university. As one of the variables is already given in the form of rank we decided to use Spearman's rank correlation coefficient $\rho$ as the measure of dependence between $r$ and $n$. The results are gathered in Table \ref{tab:all} together with the total number of papers in the given category and the statistical significance of the test.

\begin{figure}[!ht]
\begin{center}
\includegraphics[width=4.1in]{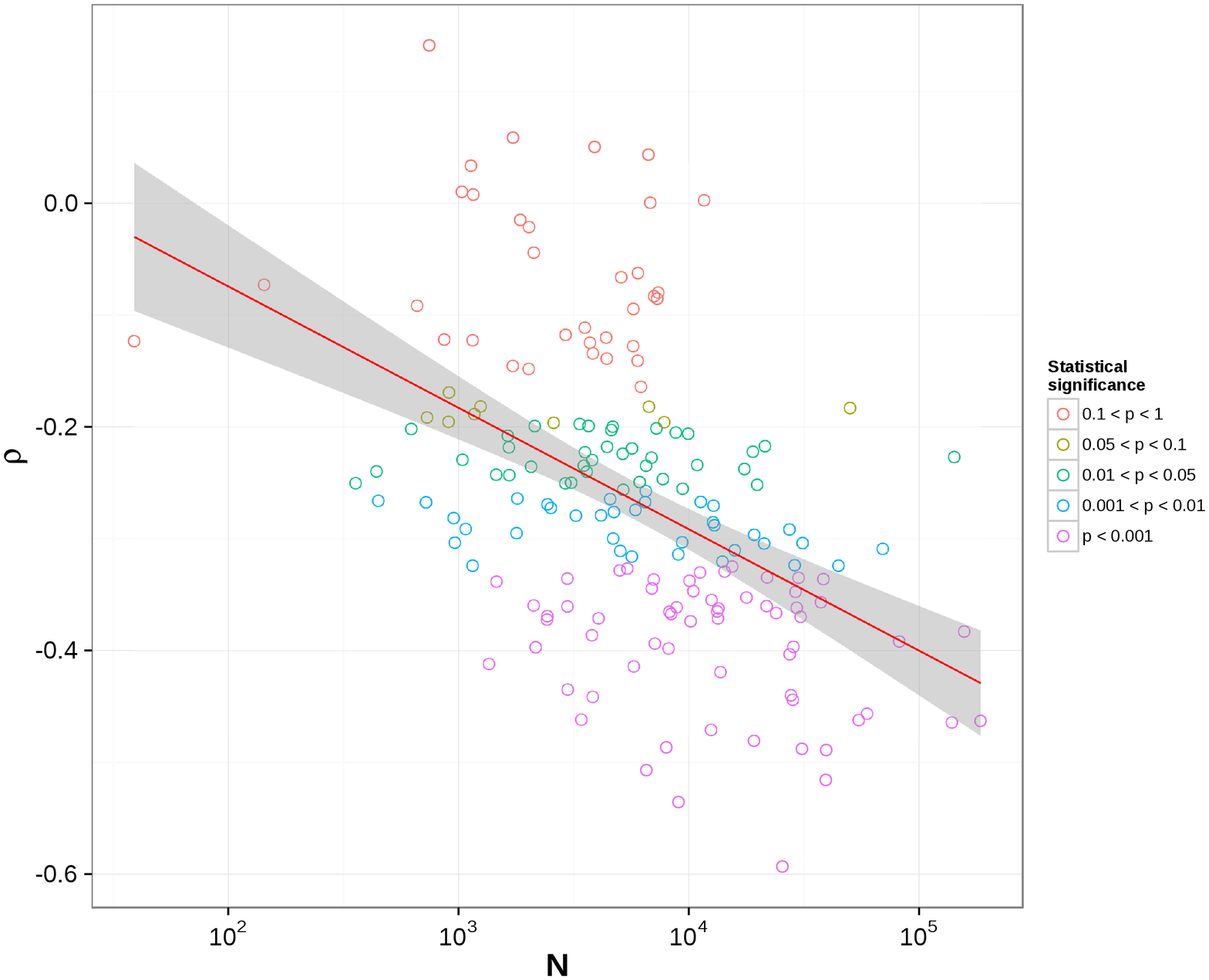}
\end{center}
\caption{{\bf Correlation vs. category size.}  }
\label{fig:rho}
\end{figure}

It is of use to examine the relation between the size of the category, measured by the number of papers $N$ belonging to it and the above mentioned correlation coefficient $\rho$. Those results are shown in Fig. \ref{fig:rho} giving the evidence that lower correlation (i.e., larger number of papers following higher rank) is in general characteristic for categories with large total number of papers. Moreover, the correlation for such categories are statistically significant.  

\subsection*{Categorical separation}
\begin{figure}[!ht]
\begin{center}
\begin{tabular}{cc}
\includegraphics[width=3in]{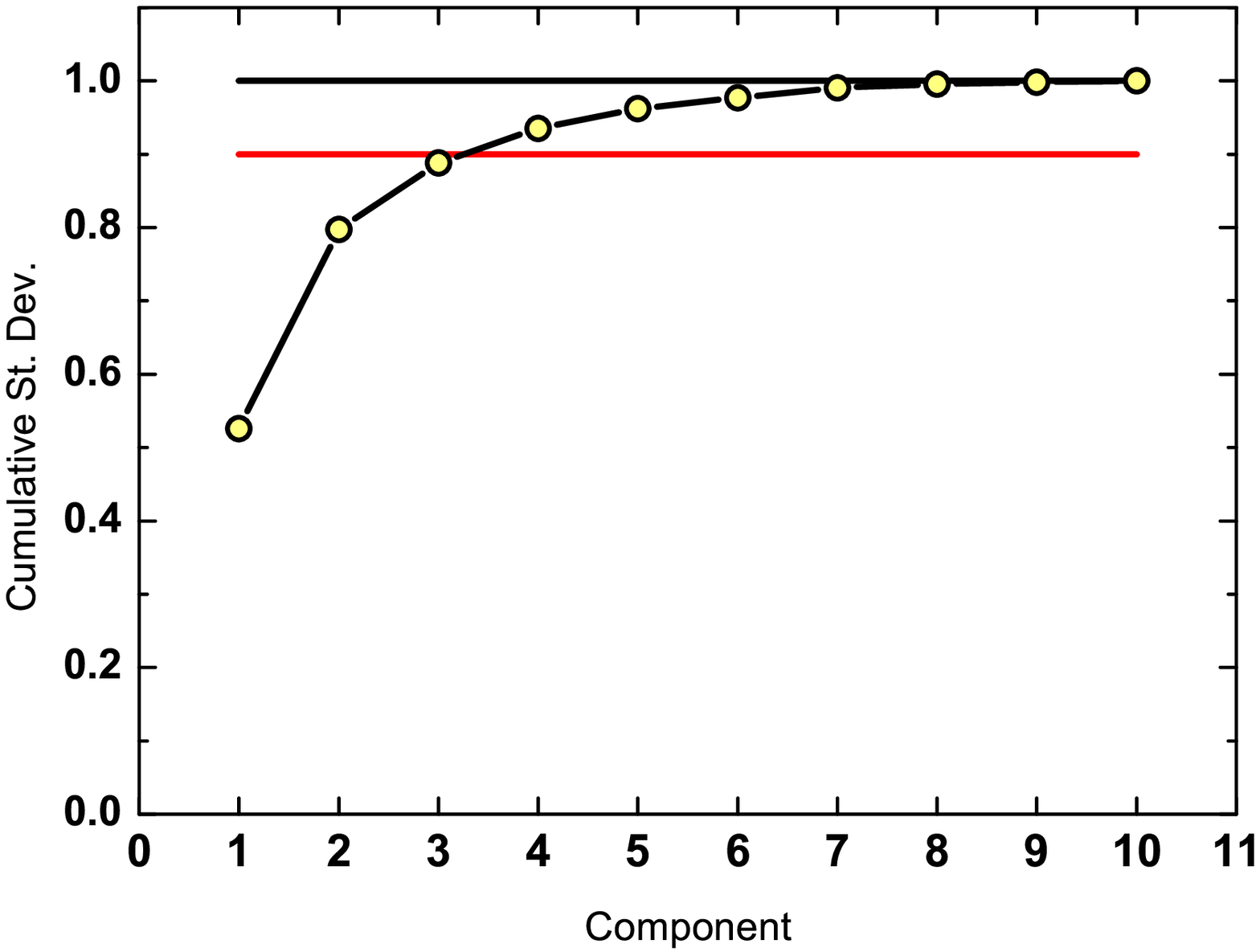} & \includegraphics[width=3in]{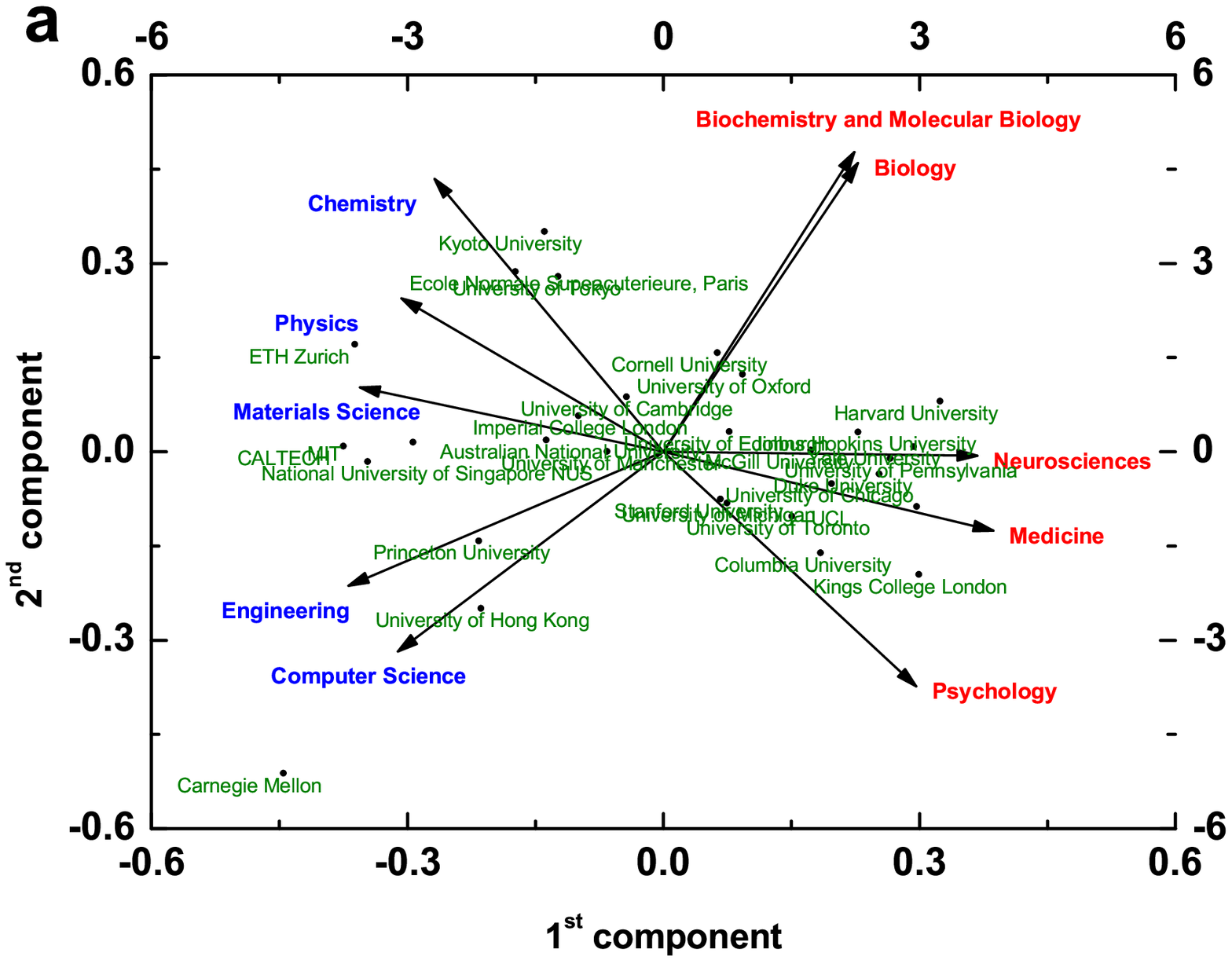}\\
\includegraphics[width=3in]{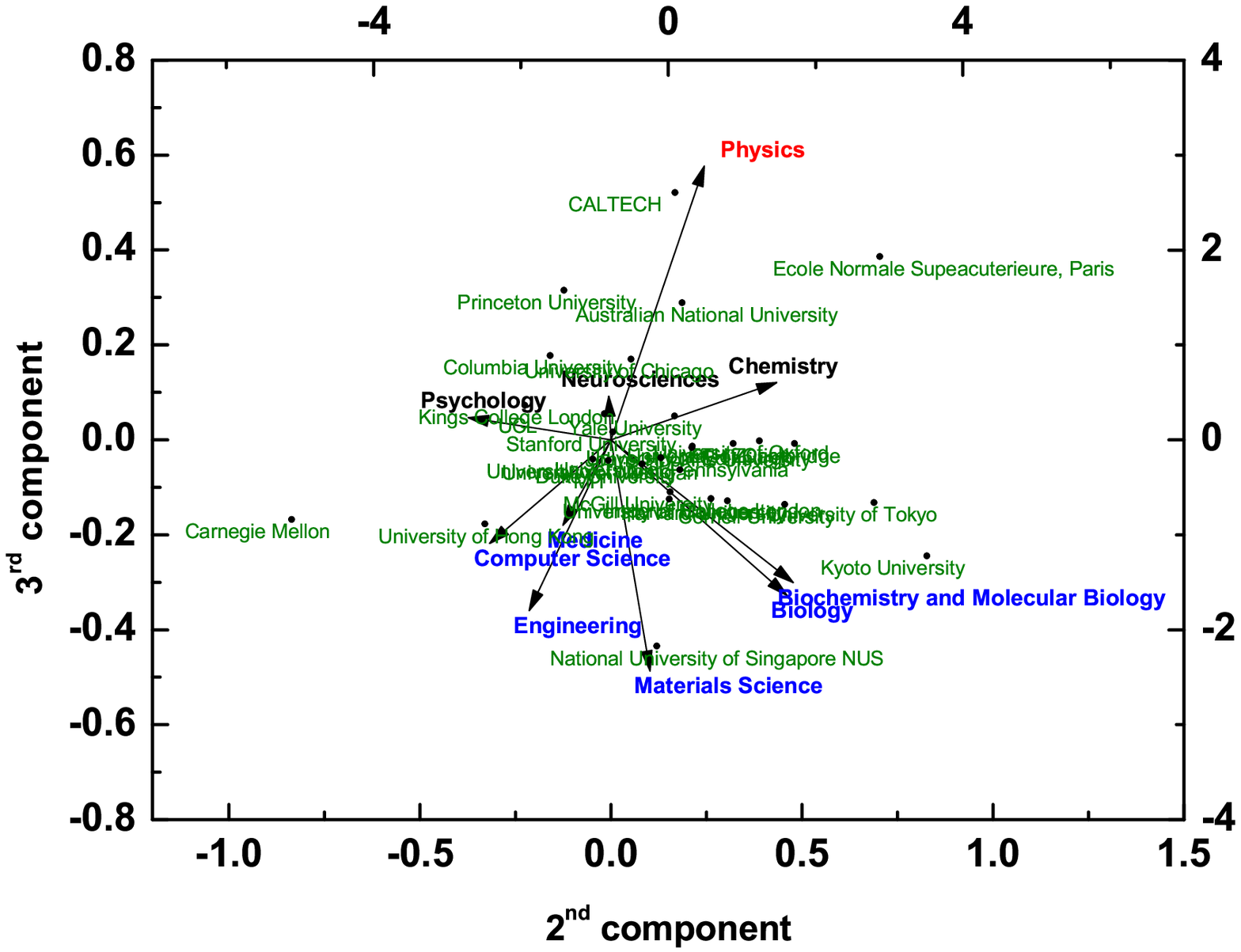} & \includegraphics[width=3in]{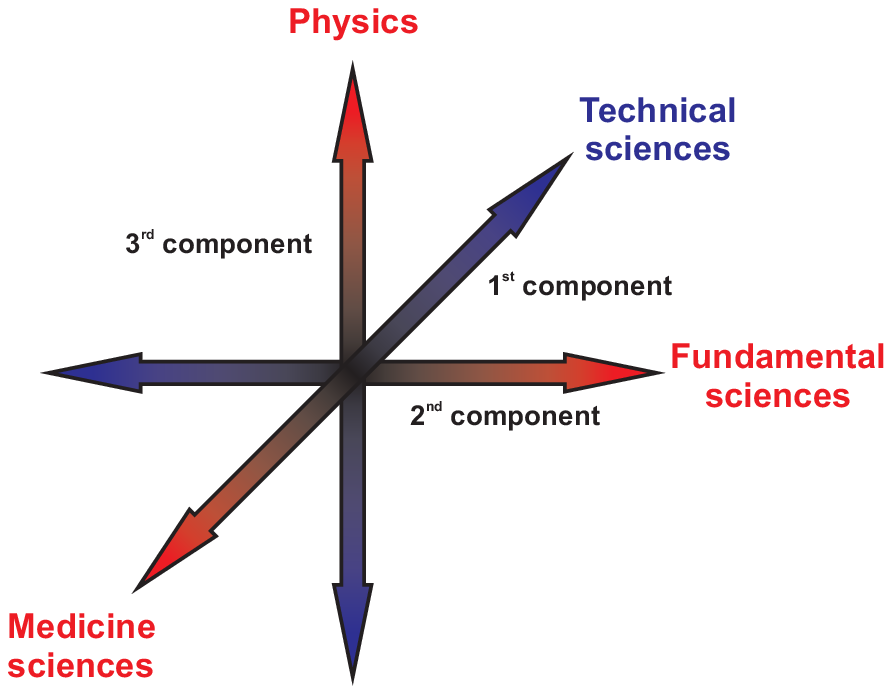}\\
\end{tabular}
\end{center}
\caption{
{\bf Principal component analysis (PCA).}}
\label{fig:pca}
\end{figure}

Here, we would like to check the hypothesis of categorical separation of Science. It is our belief, that certain categories ten to "glue together" the scientists working in them. In other words, the possibility of interdisciplinary research is not that high as one would expect it to be. In order to test this assumption we we performed the Principal Component Analysis (PCA) for 10 most prominent categories (in sense of the total number of papers). As can be seen in Fig. \ref{fig:pca}a, the first three principal components explain 90\% of variability in the data, so the analysis can be restricted just to them. Further, by plotting 2nd component vs 1st (Fig. \ref{fig:pca}b) and 3rd component vs. 2nd (Fig. \ref{fig:pca}c) we can identify the main directions of the dataset. In Fig. \ref{fig:pca}b we have biochemistry, biology, neuroscience, medicine and psychology in the positive part of x-axis while chemistry, physics, materials science, engineering and computer science are in negative part of this axis. In can thus mean that the 1st component divides the categories into technical sciences (negative values) and medicine-related ones (positive values). The 2nd component is much harder to be identified --- a rough estimate could link positive axis with {\it fundamental sciences} as we have physics, chemistry and biology. Finally there is a clear interpretation as to the 3rd component --- the only significant positive value is connected to physics.

\subsection*{Network analysis}
\begin{figure}[!ht]
\begin{center}
\includegraphics[width=2.1in]{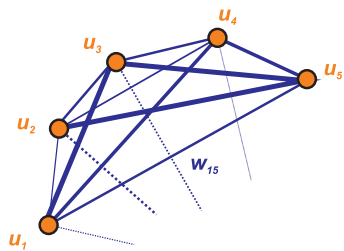}
\end{center}
\caption{{\bf Representation of the university collaboration network.}  Each node is a university and links show the connections between them. The width of each link corresponds to the number of common publications between the nodes in question.}
\label{fig:sch1}
\end{figure}

Apart from the categorical point of view we can also consider university quality by analyzing the direct connections between universities $i$ and $j$ on the basis of the collaboration matrix $\mathbf{C_{ij}}$ where the element $\mathbf{C_{ij}}$ gives the number of common publications of institutions $i$ and $j$. The principal concept of the network analysis is depicted in Fig. \ref{fig:sch1}. 

Using 100 highest ranked universities, for each of them ($u_1,u_2,...,u_{100}$) we search for its publications $p_1,p_2,...,p_{M(u1)}$. Then, if among the co-authors of $p_1$ there is any that comes from either of the universities $u_2,...,u_{100}$ a link of weight $w=1$ between those universities (e.g, $u_1$ and $u_2$) is established. The weight is increased by one each time $u_2$ is found among the following publications of $u_1$. Finally the weight of the link between nodes $u_1$ and $u_2$ is just the number of their common publications (as seen in the database).

\begin{figure}[!ht]
\begin{center}
\includegraphics[width=4in]{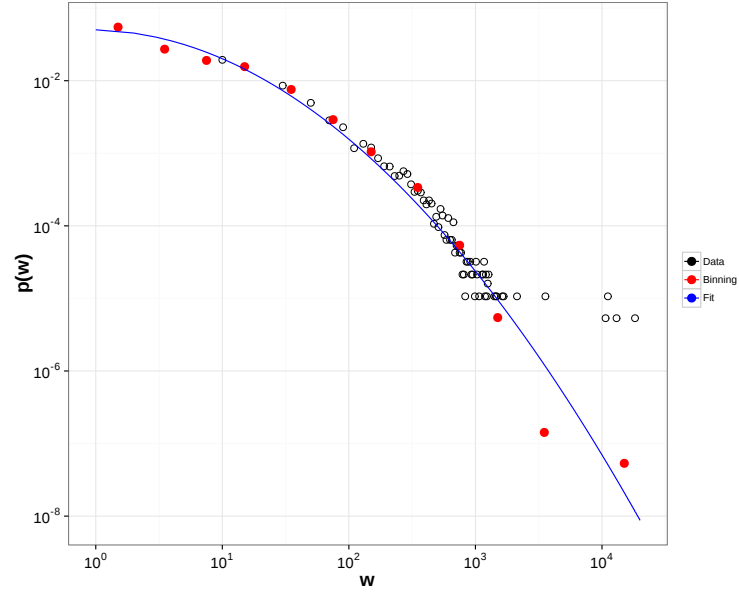}
\end{center}
\caption{{\bf Probability distribution of weights $p(w)$ in the data.}  Black circles are original data, red-filled circles are binned data (logarithmic binning) and the blue line is a fit to the binned data given by Eq. (\ref{eq:pw}).}
\label{fig:pw}
\end{figure}

\paragraph{Weights probability distribution} The first, fundamental quantity to be computed is the probability distribution of weights $p(w)$ giving the idea about the diversity of number of publications between universities. Figure \ref{fig:pw} presents $p(w)$ for raw data (black circles) as well as for the logarithmically binned ones (with the base $b=2$, red-filled circles). The plot suggests that the majority of weights can be found for w between 1 and 10 - there a plateau can be clearly seen. However, there is still a clear pattern for the remaining part even for weights as large as $w=10000$ that could be presumably fitted by a power-law function. However it is possible to fit a full-range log-normal function (red curve) 
\begin{equation}\label{eq:pw}
p(w) = \frac{1}{w \sigma \sqrt{2 \pi}} \mathrm{e}^{-\frac{ \left( \ln w - \mu \right)^2}{2 \sigma^2}}
\end{equation}
with the parameters $\mu=3.44 \pm 0.02$ and $\sigma=1.63 \pm 0.01$ (values obtained by a maximum-likelihood fitting). However, the Kolmogorov-Smirnov goodness-of-fit test accepts the hypothesis that the data points come from the distribution described by (\ref{eq:pw}) for relatively low level of significance ($\alpha=0.01$). The result is similar to this obtained in

Performing this search for consecutive universities from the ranking we obtain a fully connected network of all 100 universities with links denoting the number of common publications.

\begin{figure}
\includegraphics[width=\textwidth]{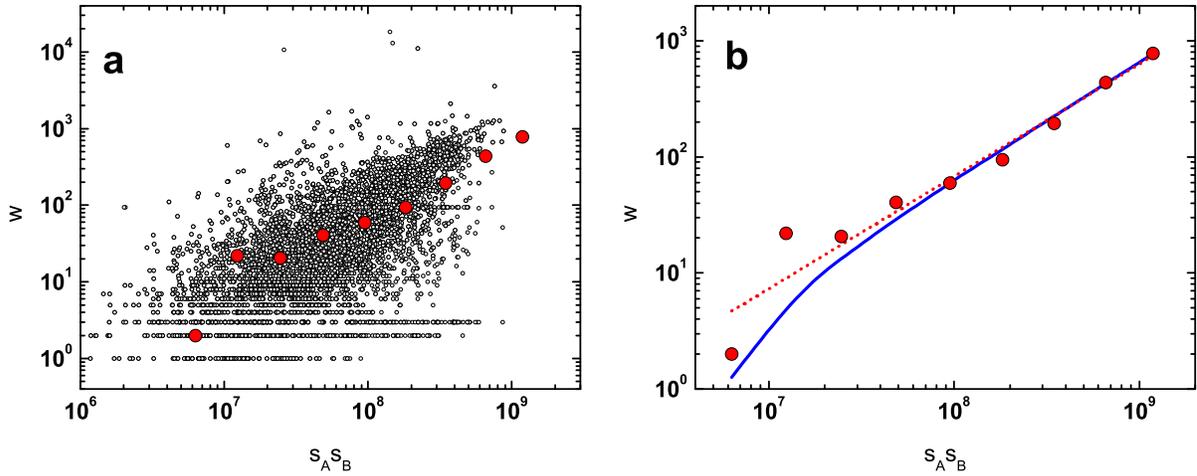}
\caption{Log-log plot of weights $w_{AB}$ versus the product of strengths $s_A s_B$. (a) Original data (black circles) and binned data (red-filled circles). (b) Log-log plot of fittings to the binned data: linear (blue solid line) and power-law (red dotted line).}
\label{fig:wsasb}
\end{figure}

\paragraph{Dependence of edge width on node strength} An interesting point of the further analysis is to test if the strength of the university, measured as the total number of its publications with other universities from the ranking influence the affinity of one university to link to another one. More precisely, we shall test what is dependence of the weight $w_{AB}$ between universities $u_A$ and $u_B$ on the product of their strengths $s_A s_B$. A log-log scatter-plot of this relation for all pairs of universities is shown in Fig. \ref{fig:wsasb}a with black circles. It brings clear evidence that the larger is the product of universities' strengths the higher is the number of common publications between them. By performing a logarithmic binning (red-filled circles) it is possible to analyze the specific form of the relation. The outcome is presented in Fig. \ref{fig:wsasb}b, where two fits are shown: a linear one (blue line, slope $a = 6.54 \times 10^{-7} \pm 0.15 \times 10^{-7}$ and negligible intercept) and power-law one (red dotted line, exponent $0.97 \pm 0.01$). The linear fitting has the $R^2$ value of 0.99 while the power-law one 0.94. Taking into account those value as well as close to 1 exponent of the power-law fitting it is reasonable to assume that the average weight between universities characterized by strengths (number of publications) $s_A$ and $s_B$ is given by the relation
\begin{equation}\label{eq:wab}
\langle w_{AB} \rangle \propto s_A s_B.
\end{equation}
Equation (\ref{eq:wab}) can serve as a kind of predictor for estimating a possible level of cooperation between two universities. Also, observed deviations from this law could indicate either a presence of outliers in a given dataset or invalid data, thus Eq. (\ref{eq:wab}) might be useful as a first-step verification procedure of the examined data.

\begin{figure}
\includegraphics[width=\textwidth]{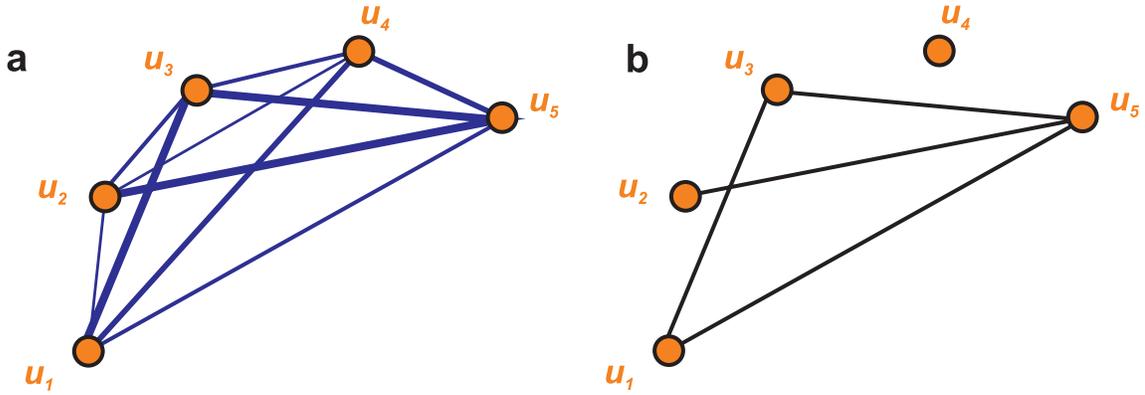}
\caption{Illustration of the weight threshold concept: (a) a weighted university network with weights proportional to the number of common publications, (b) an unweighted network constructed from the weighted one seen on panel (a) by imposing a weight threshold --- only links with weights $w > w_T$ are kept.}
\label{fig:sch2}
\end{figure}

\paragraph{Weight threshold} Following analysis will use the concept of weight threshold \cite{chmiel} depicted in Fig. \ref{fig:sch2}. Let us take the original network of 5 fully connected universities from Fig. \ref{fig:sch2}a. Let us assume now that we are interested in constructing an unweighted network that would take into account only the connections with weight higher than a certain threshold weight $w_T$ ($w > w_T$). A possible outcome of this procedure is presented in Fig. \ref{fig:sch2}b - all the links with $w < w_T$ are omitted and as a result we obtain a network where links indicate only connections between nodes (i.e., they do not bear any value). 

Using weight threshold as a parameter it is possible to obtain several unweighted networks - for each value of $w_T$ in the range $\langle w_{min}; w_{max} \rangle$ we get a different network $NT(w_T)$ whose structure is determined only by $w_T$. Then, for each of these networks it is possible to compute standard network quantities: (i) number of nodes $N$ that have a at least one link (i.e., nodes with degree $k_i=0$ are not taken into account), (ii) Number of edges (links) $E$ between the nodes,(iii) clustering coefficient $C$,  (iv) assortativity coefficient $r$ (v) entropy $S$ of node degree probability distribution and (vi) the average shortest path $\langle l \rangle$ (see Materials and Methods for details).

\begin{figure}
\includegraphics[width=\textwidth]{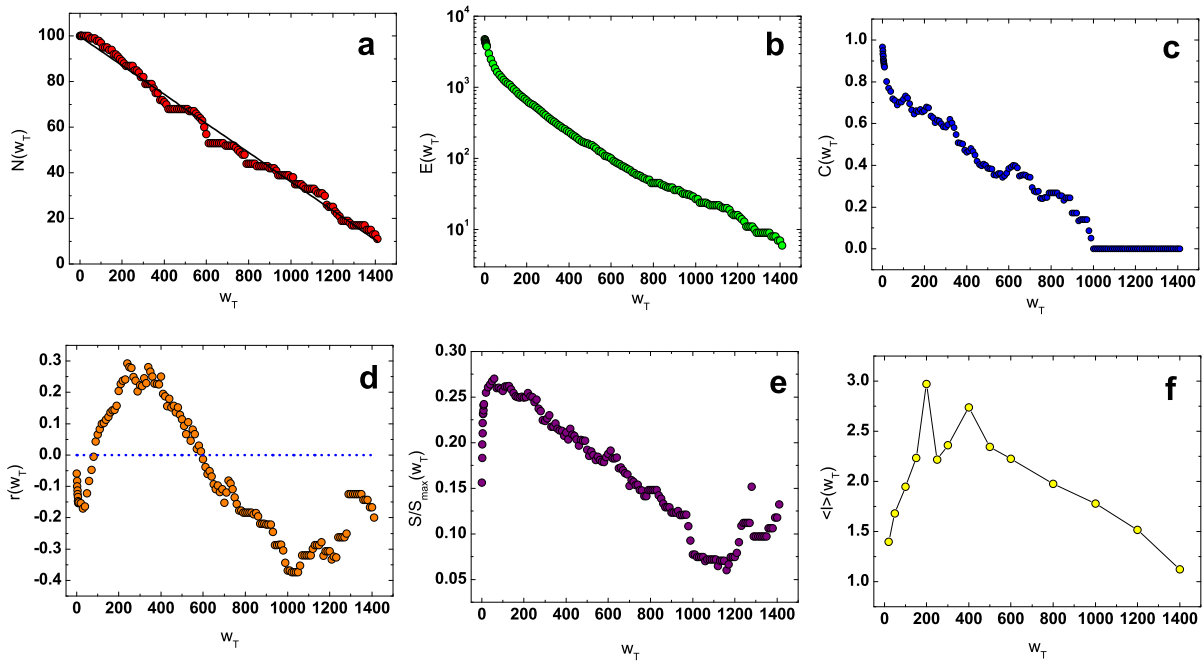}
\caption{Network observables: (a) number of nodes $N$ (b) number of edges $E$, (c) clustering coefficient $C$, (d) assortativity coefficient $r$, (e) normalized entropy and (f) average shortest path $\langle l \rangle$ as functions of weight threshold $w_T$.}
\label{fig:allw}
\end{figure}

\paragraph{Network observables as function of weight threshold} Figure \ref{fig:allw} gathers the plots of the above described network parameters as a function of $w_T$. First, as can be seen in Fig. \ref{fig:allw}a, the number of nodes $N$ is a linearly decreasing function of the weight threshold $w_T$. The number edges $E$ decreases even faster - for $w > 200$ it follows an exponential function (Fig. \ref{fig:allw}b). Similarly to $N(w_T)$ the clustering coefficient also drops down linearly with the weight threshold (Fig. \ref{fig:allw}c), although several small jumps over the trend can be seen. The most interesting is the behaviour of $r(w_T)$ shown in Figure 5d: the coefficient starts with $r<0$, while for larger thresholds it crosses $r=0$ and for $w_T$ in range $[200;400]$ it takes its maximal value. Then it once again drops down below zero reaching $r=-0.4$ for $w_T$ around 1000. Finally it increases toward zero for large $w_T$. A non-monotonic behaviour is also observed in the case of the normalized entropy $S/S_{max}$ (Fig. \ref{fig:allw}e) - here a rapid growth occurs at the very beginning (for small $w_T$), then a linear decrease happens. For large $w_T$ the normalized entropy $S/S_{max}$ again increases. The average shortest path (Fig. \ref{fig:allw}f) resembles the shape of $r$ except for the lack of growth at the end.

\paragraph{Network visualisation} The above described non-trivial behaviour of quantities $r$, $S/S_{max}$ and $\langle l \rangle$ cannot be the sole cause of the relations presented by Eqs (\ref{eq:pw}) and (\ref{eq:wab}). It seems that there has to be another phenomenon leading to such an effect.

\begin{figure}[!ht]
\begin{center}
\begin{tabular}{c}
\includegraphics[width=4.2in]{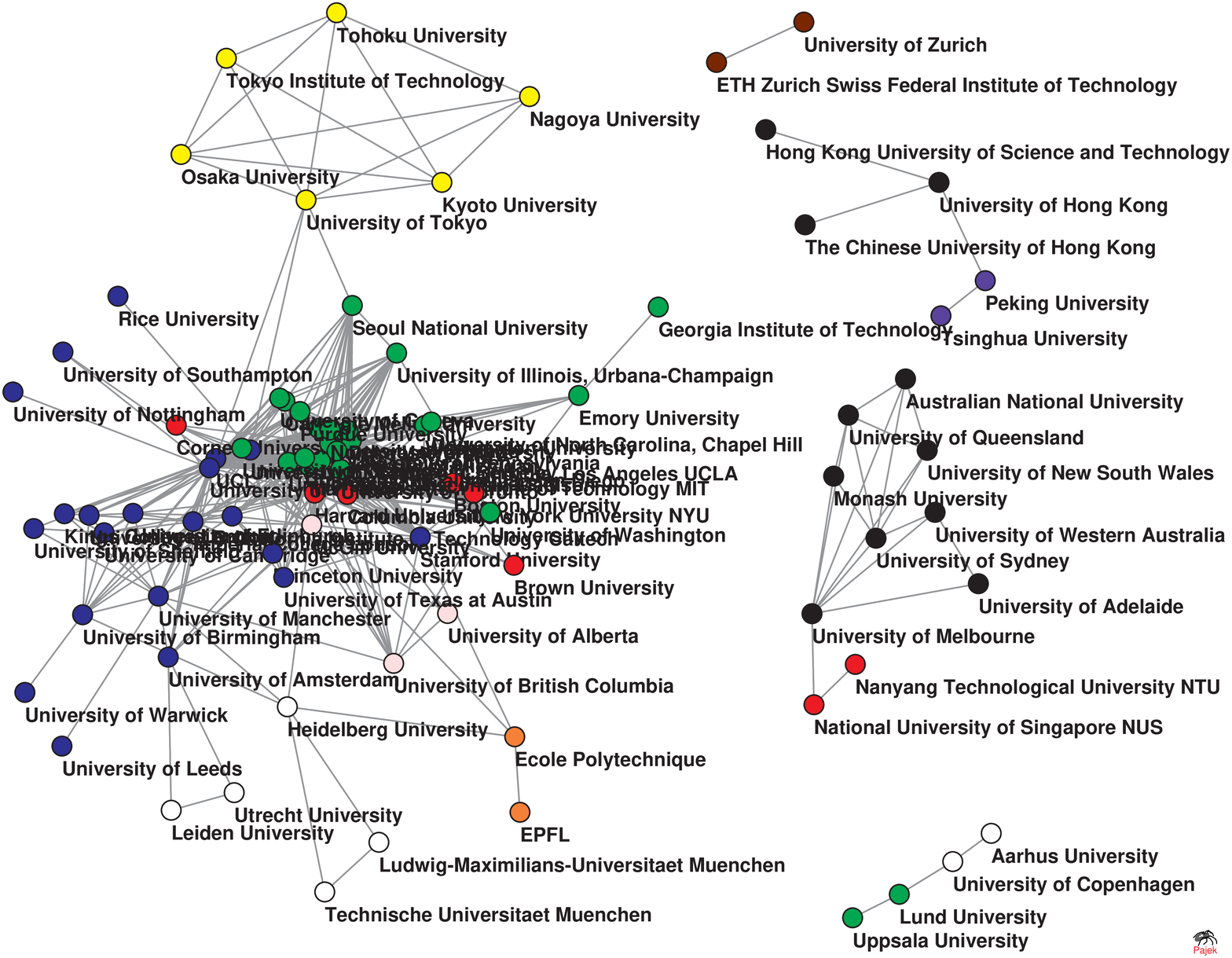}\\
\includegraphics[width=4in]{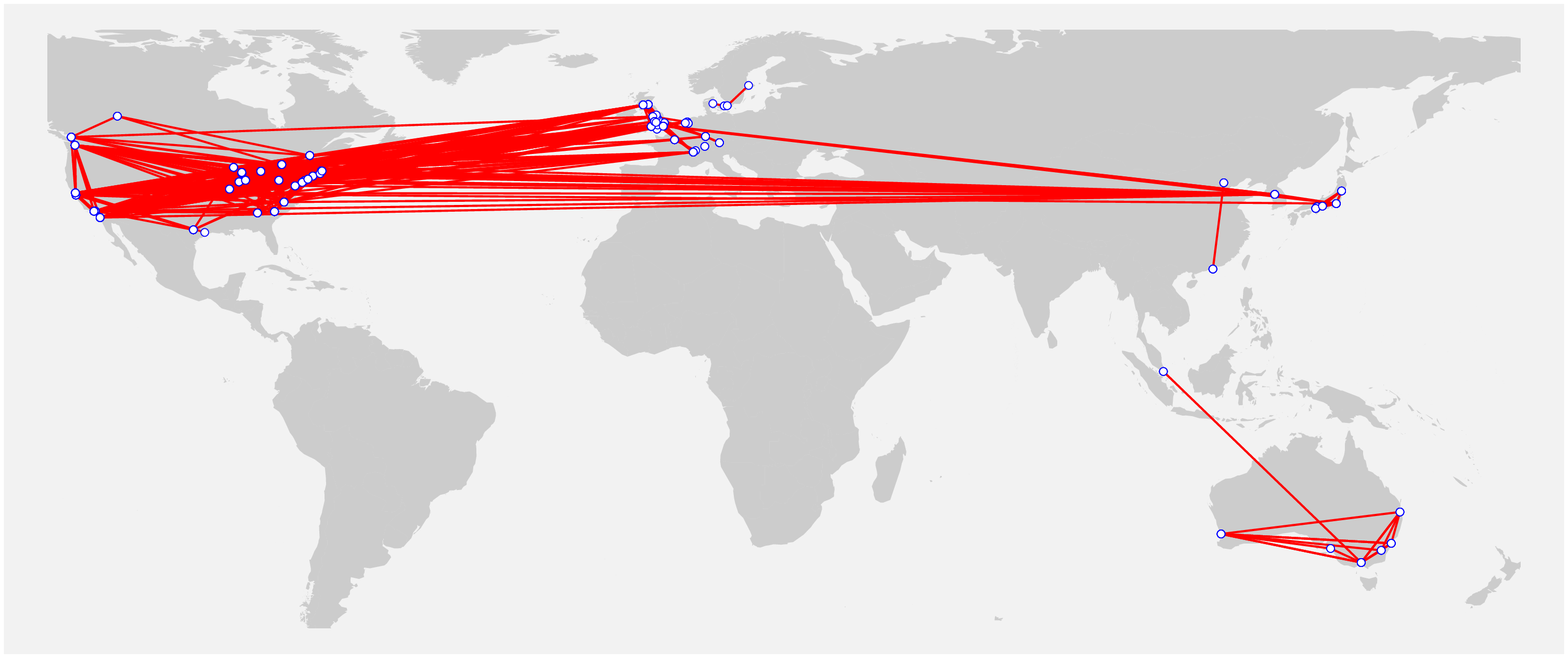}\\
\end{tabular}
\end{center}
\caption{{\bf Network for threshold $w_T=250$.}}
\label{fig:w250}
\end{figure}

\begin{figure}[!ht]
\begin{center}
\begin{tabular}{c}
\includegraphics[width=4.2in]{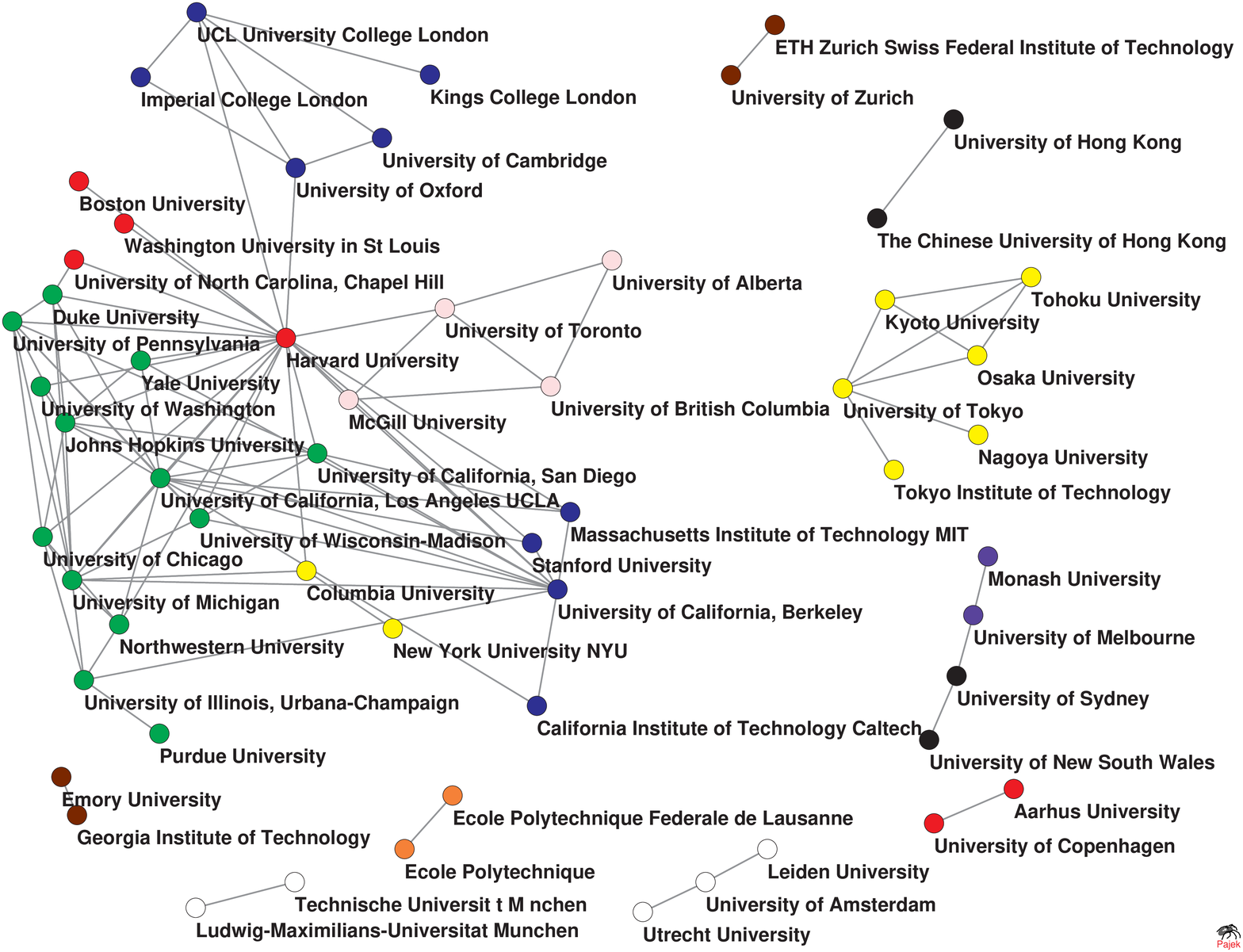}\\
\includegraphics[width=4in]{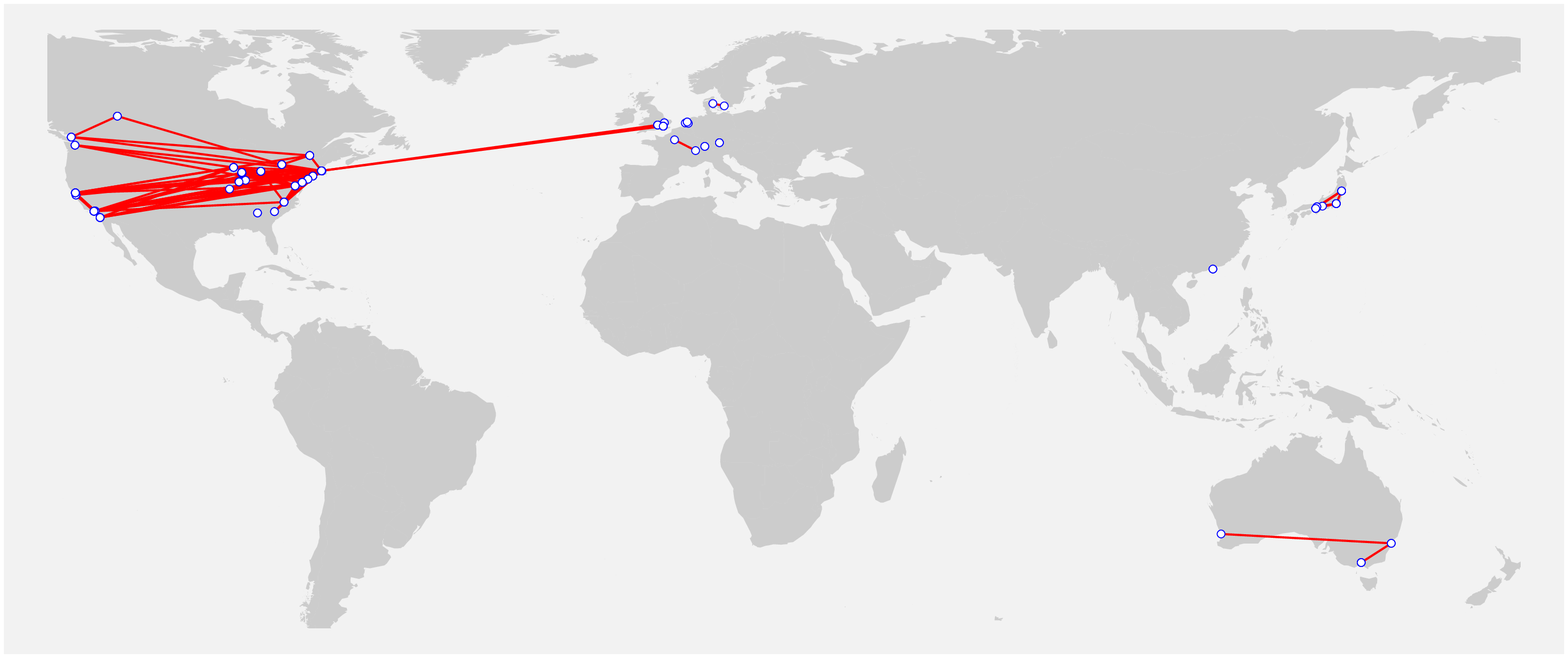}\\
\end{tabular}
\end{center}
\caption{{\bf Network for threshold $w_T=600$.}}
\label{fig:w600}
\end{figure}

\begin{figure}[!ht]
\begin{center}
\begin{tabular}{c}
\includegraphics[width=4.2in]{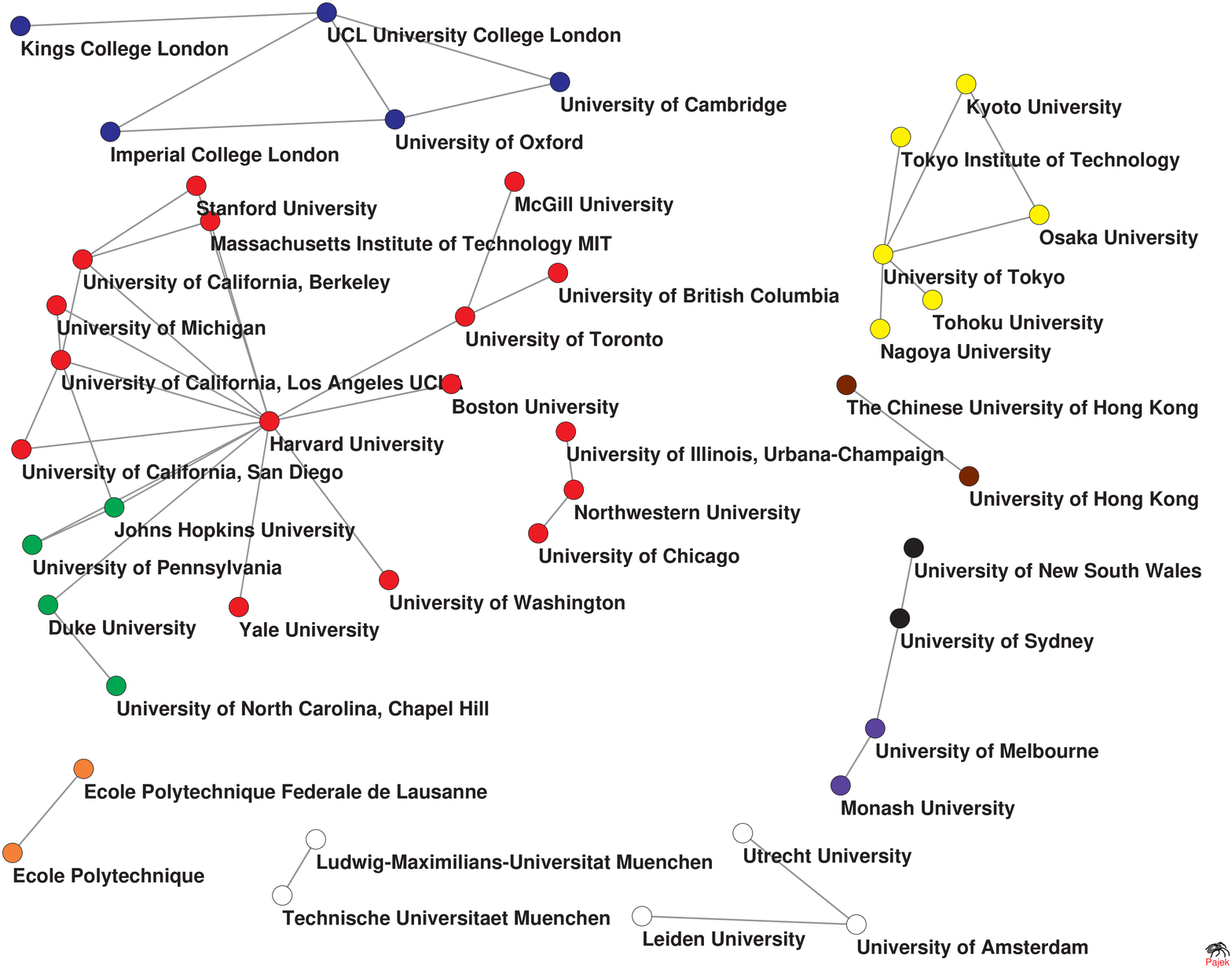}\\
\includegraphics[width=4in]{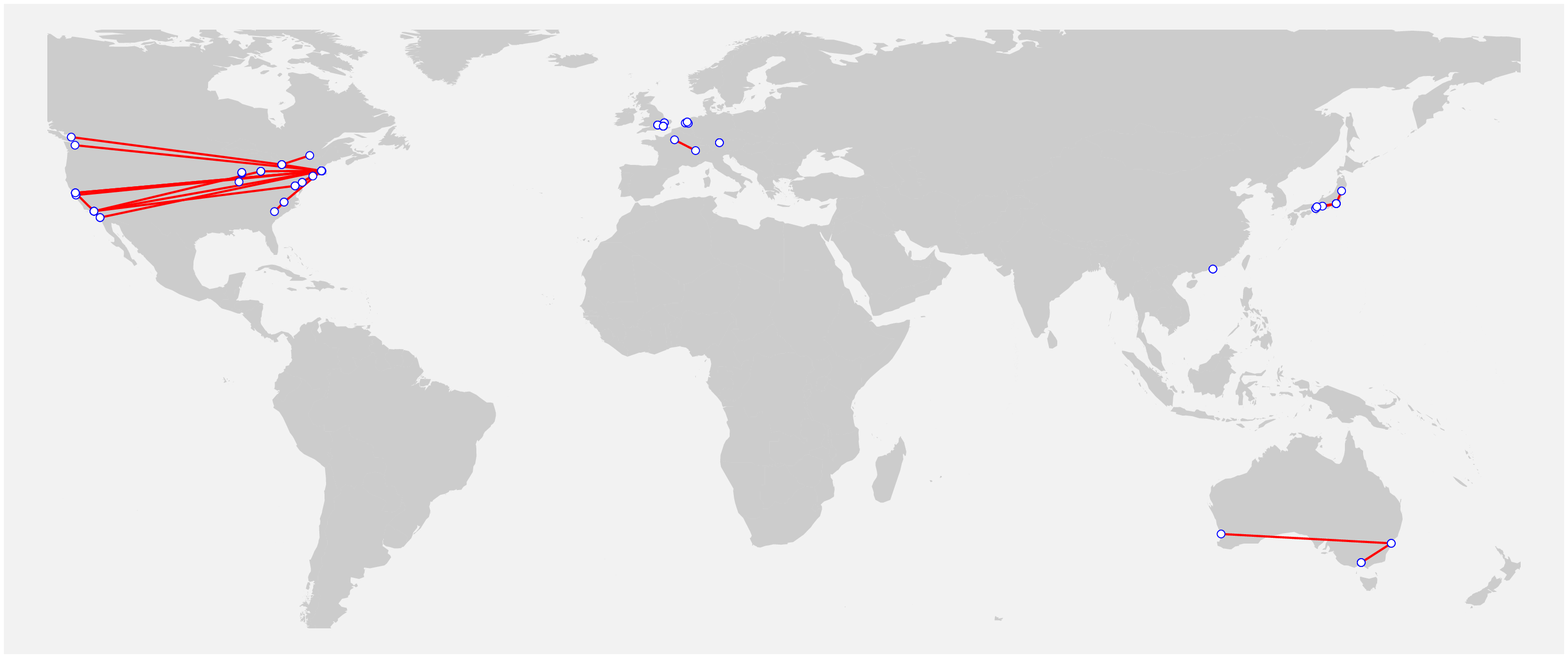}\\
\end{tabular}
\end{center}
\caption{{\bf Network for threshold $w_T=800$.}}
\label{fig:w800}
\end{figure}

\begin{figure}[!ht]
\begin{center}
\begin{tabular}{c}
\includegraphics[width=4.2in]{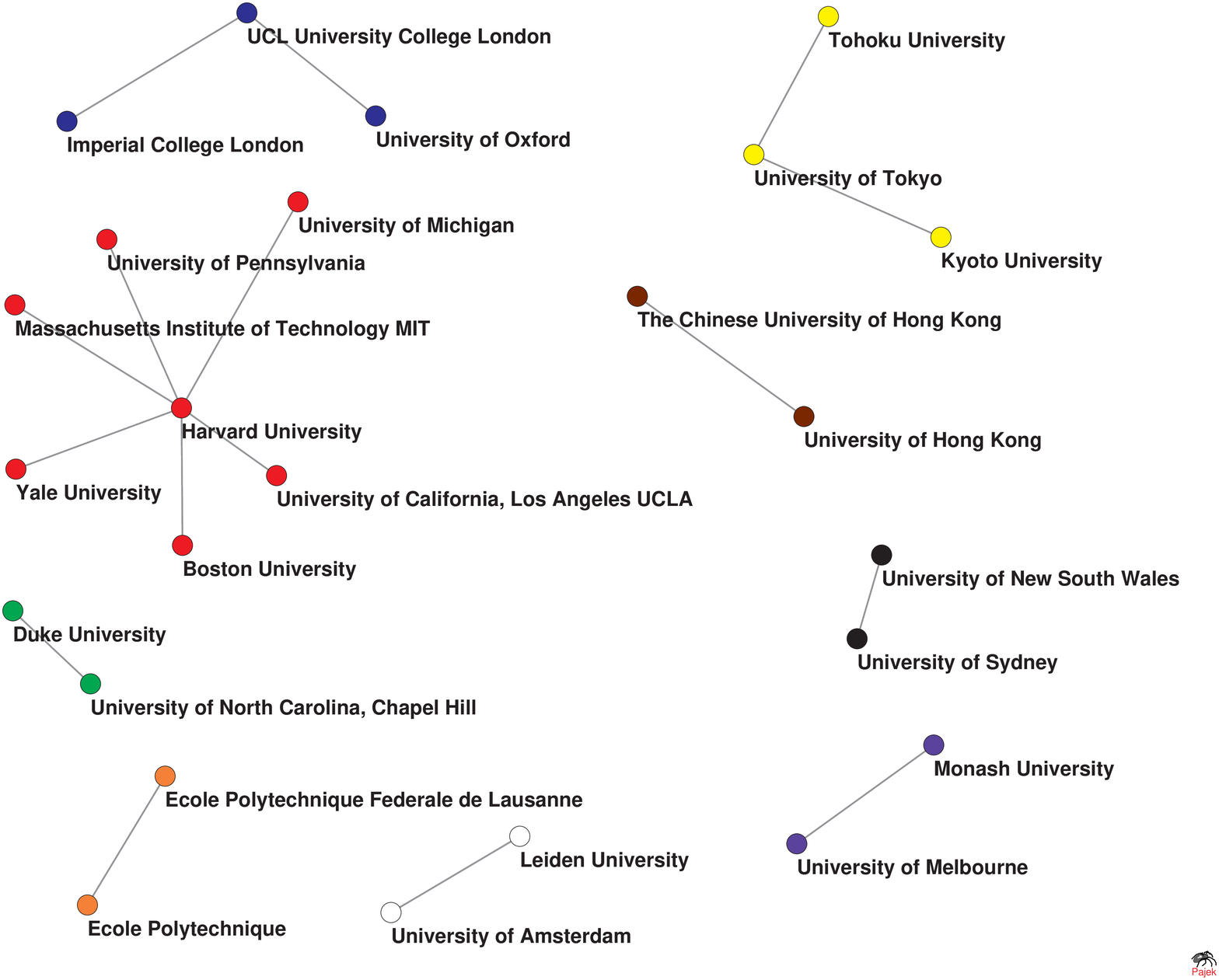}\\
\includegraphics[width=4in]{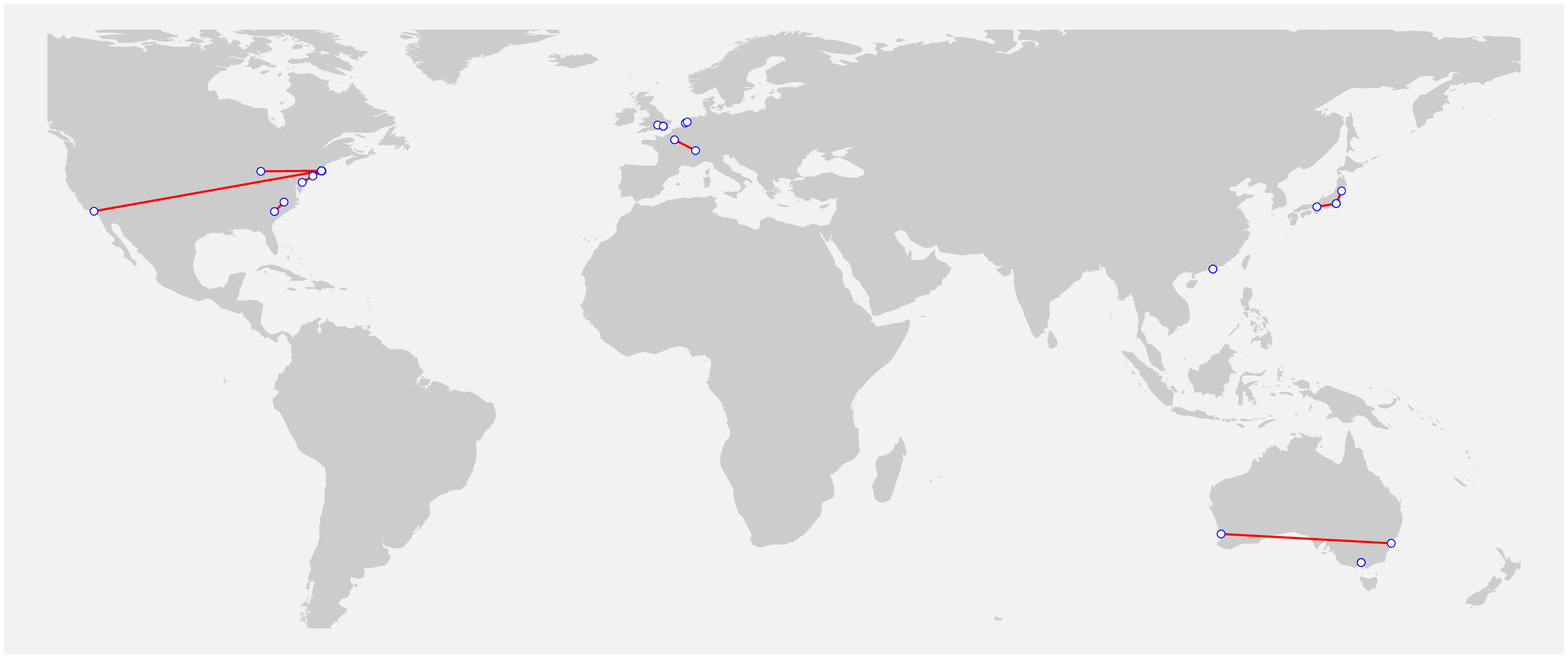}\\
\end{tabular}
\end{center}
\caption{{\bf Network for threshold $w_T=1200$.}}
\label{fig:w1200}
\end{figure}

Using Pajek\footnote{http://www.pajek.org} program as well as a tool for community detection\footnote{http://sites.google.com/site/findcommunities} it is possible to visualize connections between universities and community structure (denoted by color) for different values of $w_T$. The results for $w_T=250,600,800$ and 1200 are shown in Figs \ref{fig:w250}-\ref{fig:w1200}, providing an input for further analysis. Up to $w_T=250$ the network is percolated (it is possible to reach any node from another one); over that value a separation occurs - Chinese, Australian and Singapore, Danish and Swedish as well as Swiss universities all form separate clusters. The giant cluster is built out of American, Canadian, English, Dutch, Swiss, German and Japan universities (Fig. \ref{fig:w250}). For $w_T=600$ Dutch, Swiss, German and Japan universities are separated from the giant cluster (Fig. \ref{fig:w600}) and for $w_T=800$ also the English ones (Fig. \ref{fig:w800}). The final separation touches also American universities ($w_T=1200$, Fig. \ref{fig:w1200}).

\begin{figure}
\includegraphics[width=3in]{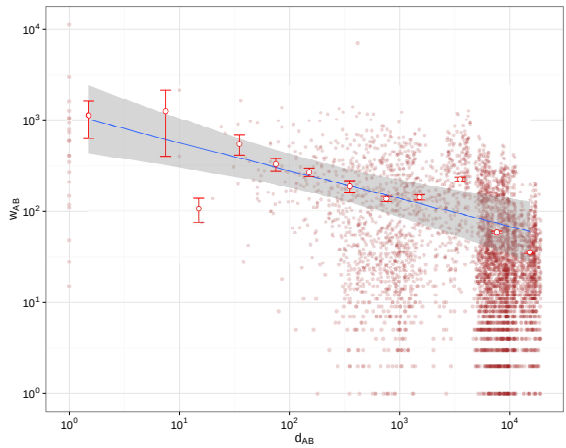}
\caption{Weight $w_{AB}$ vs geographical distance $d_{AB}$ between universities. Gray circles are row data while red-filled circles are binned data. Blue solid line is a fit (outliers are omitted during the fitting procedure).}
\label{fig:dw}
\end{figure}

It seems that the key aspect governing this kind of phenomenon is the geographical distance between the universities. In fact, Figure \ref{fig:dw} confirms this supposition: the number of publications between universities A and B is a decreasing power-law function of the geographical distance between them. The gap around $d_{AB}$=5000 is most probably caused by the presence of continents. Similar results regarding the role of geographical distance in science were obtained in \cite{rybski,pan}.

\section*{Discussion}
Our preliminary results show that even such fundamental and straightforward analysis as calculation of correlation coefficient between position of the university in the ranking and the number of papers published by its employees may reveal some non-trivial relationships. In particular, one may use it as indicator of the interest a certain scientific area gains over the years. Thus it can be possible to spot an emergence of certain trends in science and, in effect, react for example establishing a new direction of research in the university. 

Our final results show that the scientific collaboration is highly embedded in the physical  space - it seems that the key aspect that governs the number of common publications is the geographical vicinity of the universities. It is relevant even in the case of links between continents (link between Australia and Singapore). On the other hand the strength of the ties between universities is proportional to product of their total number of publications. These two relations could be used as a starting point for modeling of university collaboration.

\section*{Materials and Methods}
\par {\bf Data verification}

{\it Abbreviations}. The seemingly straightforward procedure of querying for a specific university name encounters some problems that could have a strong impact on the further results. Web of Science has a set of abbreviations commonly used for searching such as {\it Univ} for "University" or {\it Coll} for "College". Moreover it is essential to notice that one has to form a very specific query in order to get rid of severe mistakes. Table \ref{tab:univ} shows an exemplary list of the search universities together with the exact search phrase that had to be used.

{\it Ambiguity of queries}. The 'Search' field is a search key that we use to associate with the authors of the publications and it can consist of one of the operators: {\it +} which stands for {\it AND} operator in Boolean logic and {\it $|$} which stands for {\it NOT} operator in Boolean logic. These operators are used to clearly assess the origin of the publication. Table 2 shows that using just the names of universities from the list (first column) would lead in the case of number 98 to obtaining publications of both {\it Technical University in Munich} and {\it University of Munich}, instead of just the latter. To omit this problem one has to insert a query {\it Univ Munich $|$ Tech Univ Munich} that ensures achieving proper results. On the other hand for the case shown as number 78, it was not sufficient to enter {it Washington Univ}, as there are many universities with such an abbreviation; it was necessary to add {\it St. Louis} in the query text.



\par {\bf Network analysis}
{\it Clustering coefficient} $C_i$ for node $i$ is defined as the number of existing links among its nearest neighbors $e_i$ (i.e, nodes to which it has links) divided by the total number of possible links among them $k_i(k_i-1)/2$
\begin{equation}
C_i = \frac{2 e_i}{k_i(k_i - 1)}
\end{equation}
The total clustering coefficient for the whole network is calculated as the average over all $C_i$.\\
{\it Assortativity coefficient} $r$ defined by
\begin{equation}
r = \frac{ \frac{1}{E} \sum_i j_i k_i  - \left[ \frac{1}{2E} \sum_i \left(j_i + k_i \right) \right]^2}{ \frac{1}{2E} \left( \sum_i j_i^2 +  k_i^2  \right)- \left[ \frac{1}{2E} \sum_i \left(j_i + k_i \right) \right]^2}
\end{equation}
where i goes over all edges in the network. The coefficient is in the range $[-1;1]$, $r=1$ means that the highly connected nodes have the affinity to connect to other nodes with high $k_i$ while $r=-1$ happens when highly connected nodes tend to link to nodes with very low $k_i$.\\
{\it Entropy} of node degree probability distribution. It is calculated by first obtaining the degree probability distribution $p(k)$ (i.e., the probability that a randomly chosen node has exactly $k$ edges) and then evaluating the expression:
\begin{equation}
S = - \sum_{k = k_{min}}^{k = k_{max}} p(k) \ln p(k),
\end{equation}
where $k_{min}$ and $k_{max}$ are, respectively, the smallest and the largest degree in the network. For the sake of comparison we divide the obtained value of entropy by its maximal value, i.e., $S_{max}=\ln \left( k_{max}-k_{min} \right)$.
{\it Average shortest path} $\langle l \rangle$. It is calculated as the average value of shortest distance (measured in the number of steps) between all pairs of nodes $i$, $j$ in the network.

\section*{Acknowledgments}
We acknowledge support by FP7 FET Open project Dynamically Changing Complex Networks-DynaNets EU Grant Agreement Number 233847. This work has been supported by the European Union in the framework of European Social Fund through the Warsaw University of Technology Development Programme, realized by Center for Advanced Studies.


\section*{Figure Legends}

\section*{Tables}
\begin{table}[!ht]\scriptsize
\caption{\bf{Correlation coefficient in categories}}
\begin{tabular}{lrllrl}
Category & $N$ & $\rho$ & Category & $N$ & $\rho$\\
Acoustics & 5182 & -0.256* & Agricultural Economics and Policy & 440 & -0.240*\\
Agricultural Engineering & 746 & 0.141 & Agriculture & 6681 & 0.043\\
Agronomy & 2021 & -0.021 & Allergy & 3534 & -0.223*\\
Anatomy and Morphology & 1798 & -0.264** & Andrology & 448 & -0.266**\\
Anesthesiology & 4151 & -0.279** & Anthropology & 5646 & -0.316**\\
Archaeology & 2139 & -0.199* & Architecture & 12716 & -0.285**\\
Area Studies & 3791 & -0.386*** & Art & 39341 & -0.516***\\
Asian Studies & 1354 & -0.412*** & Astronomy and Astrophysics & 39504 & -0.489***\\
Automation and Control Systems & 10857 & -0.234* & Behavioral Sciences & 8221 & -0.365***\\
Biochemical Research Methods & 13696 & -0.419*** & Biochemistry and Molecular Biology & 59297 & -0.456***\\
Biodiversity Conservation & 2514 & -0.272** & Biology & 138439 & -0.464***\\
Biophysics & 13250 & -0.365*** & Biotechnology and Applied Microbiology & 17766 & -0.352***\\
Business & 10055 & -0.338*** & Cardiac and Cardiovascular Systems & 28759 & -0.324**\\
Cell and Tissue Engineering & 2418 & -0.372*** & Cell Biology & 30993 & -0.488***\\
Chemistry & 156631 & -0.383*** & Classics & 1245 & -0.182.\\
Clinical Neurology & 37408 & -0.357*** & Communication & 21374 & -0.217*\\
Computer Science & 69375 & -0.309** & Construction and Building Technology & 3531 & -0.111\\
Criminology and Penology & 1170 & -0.189. & Critical Care Medicine & 5857 & -0.274**\\
Crystallography & 3898 & 0.050 & Dance & 39 & -0.123\\
Demography & 963 & -0.303** & Dentistry & 6004 & -0.062\\
Dermatology & 7706 & -0.247* & Developmental Biology & 7983 & -0.487***\\
Ecology & 28411 & -0.397*** & Economics & 19182 & -0.481***\\
Education and Educational Research & 6702 & -0.182. & Electrochemistry & 4368 & -0.120\\
Emergency Medicine & 2906 & -0.250* & Endocrinology and Metabolism & 23900 & -0.366***\\
Energy and Fuels & 7356 & -0.080 & Engineering & 141785 & -0.227*\\
Entomology & 2120 & -0.044 & Environmental Sciences & 19185 & -0.296**\\
Environmental Studies & 4686 & -0.300** & Ergonomics & 1033 & 0.010\\
Ethics & 2116 & -0.360*** & Ethnic Studies & 729 & -0.191.\\
Evolutionary Biology & 9341 & -0.303** & Family Studies & 1784 & -0.295**\\
Film & 3826 & -0.134 & Fisheries & 1724 & 0.059\\
Folklore & 143 & -0.073 & Food Science and Technology & 6792 & 0.000\\
Forestry & 2014 & -0.148 & Gastroenterology and Hepatology & 15405 & -0.325***\\
Genetics and Heredity & 27733 & -0.440*** & Geochemistry and Geophysics & 14279 & -0.329***\\
Geography & 7055 & -0.083 & Geology & 2912 & -0.118\\
Geosciences & 17401 & -0.238* & Geriatrics and Gerontology & 5761 & -0.414***\\
Gerontology & 8154 & -0.398*** & Health Care Sciences and Services & 10164 & -0.374***\\
Health Policy and Services & 6895 & -0.344*** & Hematology & 29896 & -0.335***\\
History and Philosophy Of Science & 3411 & -0.462*** & History Of Social Sciences & 1460 & -0.338***\\
Horticulture & 1157 & 0.008 & Hospitality & 1131 & 0.034\\
Humanities & 5006 & -0.328*** & Imaging Science and Photographic Technology & 5033 & -0.311**\\
Immunology & 27363 & -0.403*** & Industrial Relations and Labor & 1039 & -0.229*\\
Infectious Diseases & 13355 & -0.371*** & Information Science and Library Science & 3599 & -0.240*\\
Instruments and Instrumentation & 8767 & -0.205* & Integrative and Complementary Medicine & 904 & -0.195.\\
International Relations & 2966 & -0.360*** & Language and Linguistics & 3799 & -0.230*\\
\end{tabular}
\begin{flushleft}\end{flushleft}
\label{tab:all}
\end{table}

\begin{table}[!ht]\scriptsize
\caption{\bf{Correlation coefficient in categories (ctnd)}}
\begin{tabular}{lrllrl}
Category & $N$ & $\rho$ & Category & $N$ & $\rho$\\
Law & 4049 & -0.371*** & Limnology & 1718 & -0.145\\
Linguistics & 6109 & -0.249* & Literary Reviews & 951 & -0.281**\\
Literary Theory and Criticism & 722 & -0.267** & Literature & 6879 & -0.227*\\
Management & 12783 & -0.270** & Marine and Freshwater Biology & 5079 & -0.066\\
Materials Science & 50040 & -0.183. & Mathematical and Computational Biology & 6544 & -0.507***\\
Mathematics & 28228 & -0.444*** & Mechanics & 11252 & -0.267**\\
Medical Ethics & 1150 & -0.324** & Medical Informatics & 2975 & -0.435***\\
Medical Laboratory Technology & 2429 & -0.269** & Medicine & 81833 & -0.392***\\
Medieval and Renaissance Studies & 1073 & -0.291** & Metallurgy and Metallurgical Engineering & 6186 & -0.164\\
Meteorology and Atmospheric Sciences & 10451 & -0.347*** & Microbiology & 30534 & -0.370***\\
Microscopy & 908 & -0.169. & Mineralogy & 2064 & -0.236*\\
Mining and Mineral Processing & 1150 & -0.122 & Multidisciplinary Sciences & 25490 & -0.593***\\
Music & 1456 & -0.243* & Mycology & 866 & -0.122\\
Nanoscience and Nanotechnology & 19773 & -0.252* & Neuroimaging & 3821 & -0.441***\\
Neurosciences & 54583 & -0.462*** & Nuclear Science and Technology & 5157 & -0.224*\\
Nursing & 4410 & -0.218* & Nutrition and Dietetics & 7805 & -0.196.\\
Obstetrics and Gynecology & 13415 & -0.362*** & Oceanography & 4667 & -0.200*\\
Oncology & 38328 & -0.336*** & Operations Research and Management Science & 6452 & -0.267**\\
Ophthalmology & 8838 & -0.361*** & Optics & 21195 & -0.304**\\
Ornithology & 660 & -0.092 & Orthopedics & 6515 & -0.235*\\
Otorhinolaryngology & 3495 & -0.235* & Paleontology & 3080 & -0.250*\\
Parasitology & 3229 & -0.279** & Pathology & 12537 & -0.355***\\
Pediatrics & 15804 & -0.310** & Peripheral Vascular Disease & 21836 & -0.335***\\
Pharmacology and Pharmacy & 27303 & -0.292** & Philosophy & 7109 & -0.394***\\
Physics & 184647 & -0.463*** & Physiology & 13971 & -0.320**\\
Planning and Development & 2427 & -0.369*** & Plant Sciences & 11636 & 0.003\\
Poetry & 357 & -0.250* & Political Science & 7026 & -0.336***\\
Polymer Science & 7222 & -0.201* & Psychiatry & 29348 & -0.362***\\
Psychology & 44591 & -0.324** & Public & 29017 & -0.347***\\
Public Administration & 1635 & -0.208* & Radiology & 21722 & -0.360***\\
Rehabilitation & 5722 & -0.128 & Religion & 3665 & -0.199*\\
Remote Sensing & 2585 & -0.196. & Reproductive Biology & 6494 & -0.257**\\
Respiratory System & 11167 & -0.330*** & Rheumatology & 9382 & -0.255*\\
Robotics & 4601 & -0.203* & Social Issues & 2159 & -0.397***\\
Social Sciences & 12472 & -0.471*** & Social Work & 1659 & -0.243*\\
Sociology & 5402 & -0.327*** & Soil Science & 1852 & -0.015\\
Spectroscopy & 4722 & -0.276** & Sport Sciences & 5988 & -0.141\\
Statistics and Probability & 9024 & -0.536*** & Substance Abuse & 4548 & -0.264**\\
Surgery & 31145 & -0.304** & Telecommunications & 18936 & -0.222*\\
Theater & 623 & -0.202* & Thermodynamics & 3353 & -0.197*\\
Toxicology & 5656 & -0.219* & Transplantation & 8991 & -0.314**\\
Transportation & 4395 & -0.139 & Transportation Science and Technology & 3717 & -0.125\\
Tropical Medicine & 2965 & -0.336*** & Urban Studies & 1652 & -0.218*\\
Urology and Nephrology & 12891 & -0.288** & Veterinary Sciences & 7279 & -0.085\\
Virology & 8370 & -0.367*** & Water Resources & 5735 & -0.094\\
Zoology & 9917 & -0.206* & & & \\
\end{tabular}
\begin{flushleft}\end{flushleft}
\label{tab:all1}
\end{table}

\begin{table}[!ht]
\caption{\bf{University names and search queries}}
\begin{tabular}{cll}
Rank & University & Search query \\
1 & Harvard University & Harvard Univ\\
2 & University of Cambridge & Univ Cambridge\\
4 & UCL University College London & UCL\\
10 & California Institute of Technology & Caltech\\
73 & Washington University in St. Louis & Washington Univ + St Louis\\
98 & Ludwig-Maximilians-Universit\"{a}t M\"{u}nchen & Univ Munich $|$ Tech Univ Munich\\
\end{tabular}
\begin{flushleft}\end{flushleft}
\label{tab:univ}
\end{table}

\end{document}